# GNNTAL:A Novel Model for Identifying Critical Nodes in Complex Networks

Hao Wang[1], Ting Luo[1], Shuang-ping Yang[1], Ming Jing[2], Jian Wang[3,*] and Na Zhao[1,*]

1：Key Laboratory in Software Engineering of Yunnan Province, Yunnan University, Kunming 650091, China

2：School of Artificial Intelligence & Information Engineering, West Yunnan University, Lincang 677000, China

3：Faculty of Information Engineering and Automation, Kunming University of Science and Technology, Kunming 650504, China

**Correspondence**：Jian Wang:jianwang@kust.edu.cn; and Na Zhao:zhaonayx@126.com

**Abstract**：Identification of critical nodes is a prominent topic in the study of complex networks. Numerous methods have been proposed, yet most exhibit inherent limitations. Traditional approaches primarily analyze specific structural features of the network; however, node influence is typically the result of a combination of multiple factors. Machine learning-based methods struggle to effectively represent the complex characteristics of network structures through suitable embedding techniques and require substantial data for training, rendering them prohibitively costly for large-scale networks. To address these challenges, this paper presents an active learning model based on GraphSAGE and Transformer, named GNNTAL. This model is initially pre-trained on random or synthetic networks and subsequently fine-tuned on real-world networks by selecting a few representative nodes using K-Means clustering and uncertainty sampling. This approach offers two main advantages: (1) it significantly reduces training costs; (2) it simultaneously incorporates both local and global features. A series of comparative experiments conducted on twelve real-world networks demonstrate that GNNTAL achieves superior performance. Additionally, this paper proposes an influence maximization method based on the predictions of the GNNTAL model, which achieves optimal performance without the need for complex computations. Finally, the paper analyses certain limitations of the GNNTAL model and suggests potential solutions.

**Keywords**：Complex Network，Active Learning，Critical Nodes，Influence Maximization

# Introduction

Complex networks are graph structures composed of nodes and edges, commonly employed to describe and analyze relationships and interactions within various complex systems in the real world, such as power grids[1,2], disease transmission networks[3,4], and rumor propagation network[5,6]. Over the past decade, complex network analysis has found extensive applications across numerous fields[7–12], with the identification of critical nodes emerging as one of the prominent research topics. Due to the heterogeneity of nodes, a small number of nodes within a network can exert a profound impact on the network's structure and function[13]. For instance, the failure of bridge nodes connecting different communities can cause network fragmentation, thereby impeding the dissemination of information within the network. Similarly, the failure of nodes positioned on the shortest or maximum flow paths can reduce the efficiency of information propagation across the network. Consequently, the critical node identification problem aims to pinpoint these nodes that significantly influence the overall structure and functionality of the network.

The problem of identifying critical nodes can be categorized into two types. The first type involves assessing the influence of individual nodes within a network and ranking them accordingly to identify critical nodes based on their influence. The second type focuses on finding a set of seed nodes that can maximize the spread of influence across the network under a given information propagation model; this type of problem is also known as the influence maximization problem[14–16]. Typically, we might select the top-ranked nodes based on their assessed influence to form the seed nodes. However, this approach presents new challenges. Firstly, accurately assessing the influence of each node in large-scale or even ultra-large-scale networks is difficult. Secondly, high-influence nodes in real networks often cluster together, forming what is known as the "rich-club" effect. This clustering causes the influence areas of high-impact nodes to overlap, resulting in information spreading only locally rather than throughout the entire network.

Existing methods for identifying critical nodes primarily analyze the statistical features of network structures[17]. These methods can be categorized into local centrality, global centrality, and semi-local centrality[18]. Local centrality identifies critical nodes based on local structural features of the network, with representative methods including SNC, Spon, SC, and the H-index[13,19–23]. Global centrality, on the other hand, identifies critical nodes based on the overall structural features of the network, with prominent methods such as K-Shell, KSIF, and PageRank[24–29]. Semi-local centrality aims to combine the advantages of both local and global centrality measures. These methods consider structural features over a broader range when evaluating node influence, but do not extend their scope to the entire network. Examples of semi-local centrality methods include Collective Influence, LID, and LGM[15,30–33]. Methods based on structural features have achieved significant success in the past; however, they often heavily rely on structural characteristics and exhibit poor generalization. Another category of critical node identification methods treats individual nodes as seed nodes and conducts simulation experiments on specific information propagation models to observe the spread of influence of these nodes. This approach enables precise assessment of a node's influence. However, propagation models typically possess probabilistic properties, making the cost of running simulation analyses on large-scale networks prohibitively high.

In recent years, the rapid advancement of machine learning and deep learning has introduced new approaches to the identification of critical nodes. Some researchers have attempted to obtain embedding vectors of network structures through embedding methods or customized features, and subsequently train machine learning models[34,35]. Following this, the emergence of Graph Neural Networks[36] (GNNs) and Convolutional Neural Networks[37] (CNNs) has broadened the horizon for researchers, who have started to explore the application of deep learning and reinforcement learning techniques to the study of complex networks[16,38–40]. While these research paradigms are highly effective, they also exhibit certain limitations. Firstly, the structural characteristics of complex networks are intricate and challenging to represent using simple methods. Secondly, deep learning methods are often based on GNNs and CNNs, which focus excessively on local features and fail to assess node influence from a global perspective[40]. Additionally, deep learning methods require large datasets for training, yet it is difficult to obtain large-scale, especially large-scale real network simulation data. This undoubtedly restricts the development potential of deep learning methods.

Recently, attention mechanism-based models like Transformer[41] and its derivatives, such as BERT[42] and GPT[43], have achieved remarkable success in the field of artificial intelligence. In this context, we have designed a critical node identification model based on GraphSAGE[44] and Transformer[41]. This model is initially pre-trained on small, artificially generated random networks and then fine-tuned on real networks using an active learning strategy based on K-Means clustering and uncertainty sampling. The advantages of this approach are as follows:

(1)Pre-training on artificial networks and fine-tuning with a small number of nodes from real networks significantly reduces training costs. This method eliminates the need for extensive real network data and numerous simulation experiments.

(2)The LSTM aggregator in GraphSAGE effectively captures complex local features within a node's neighborhood, while Transformer can capture long-range dependencies and global information among nodes. The combination of these two models allows for the simultaneous capture of both local and global features, achieving multi-scale feature fusion.

To validate the performance of GNNTAL, we conducted a series of comparative experiments. The results demonstrate the superior performance of GNNTAL. Additionally, based on the predictions of the GNNTAL model, we propose a low-complexity greedy strategy to address the influence maximization problem in complex networks. This strategy avoids redundant coverage of the influence areas of multiple critical nodes, thereby maximizing collective influence. Experimental results indicate that this strategy achieves good performance with relatively low computational complexity.

## Model

The overall framework of GNNTAL can be divided into three parts, as shown in Figure 01:

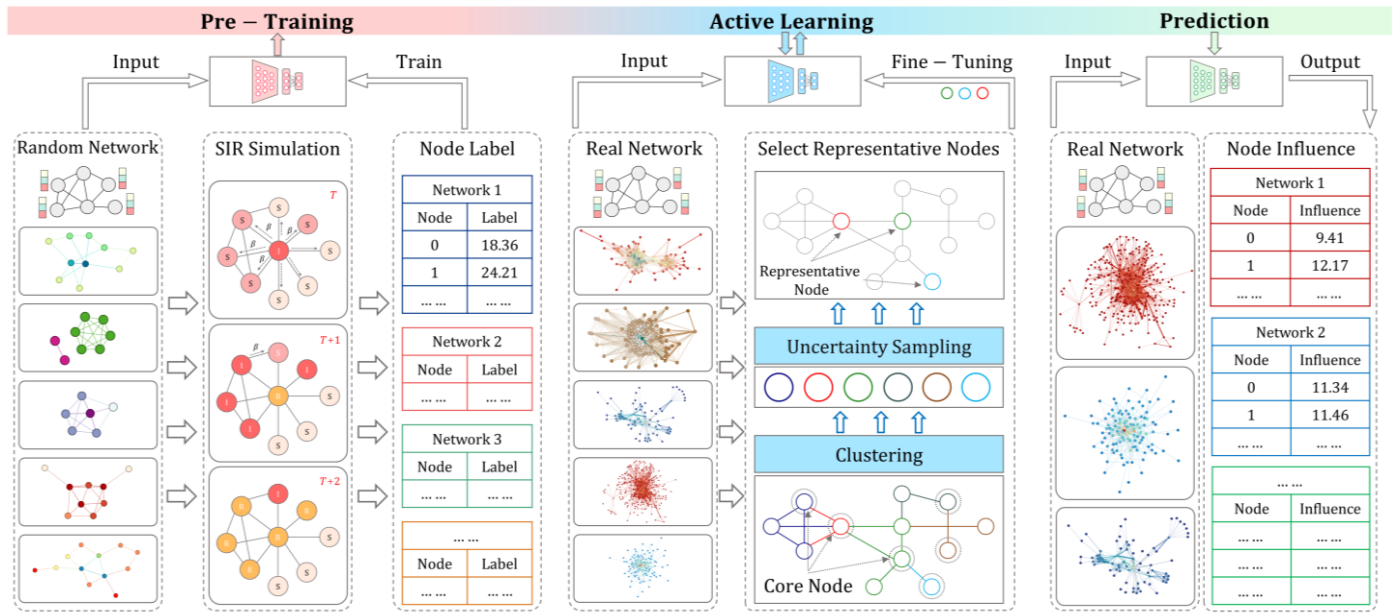

**Figure 01: Overall Framework of the GNNTAL Model** The overall framework of GNNTAL comprises three parts: (1) a node influence prediction model based on GraphSAGE and Transformer, pre-trained on artificial networks; (2) active learning conducted on real networks using K-Means clustering and uncertainty sampling; and (3) influence prediction for each node within the real networks.

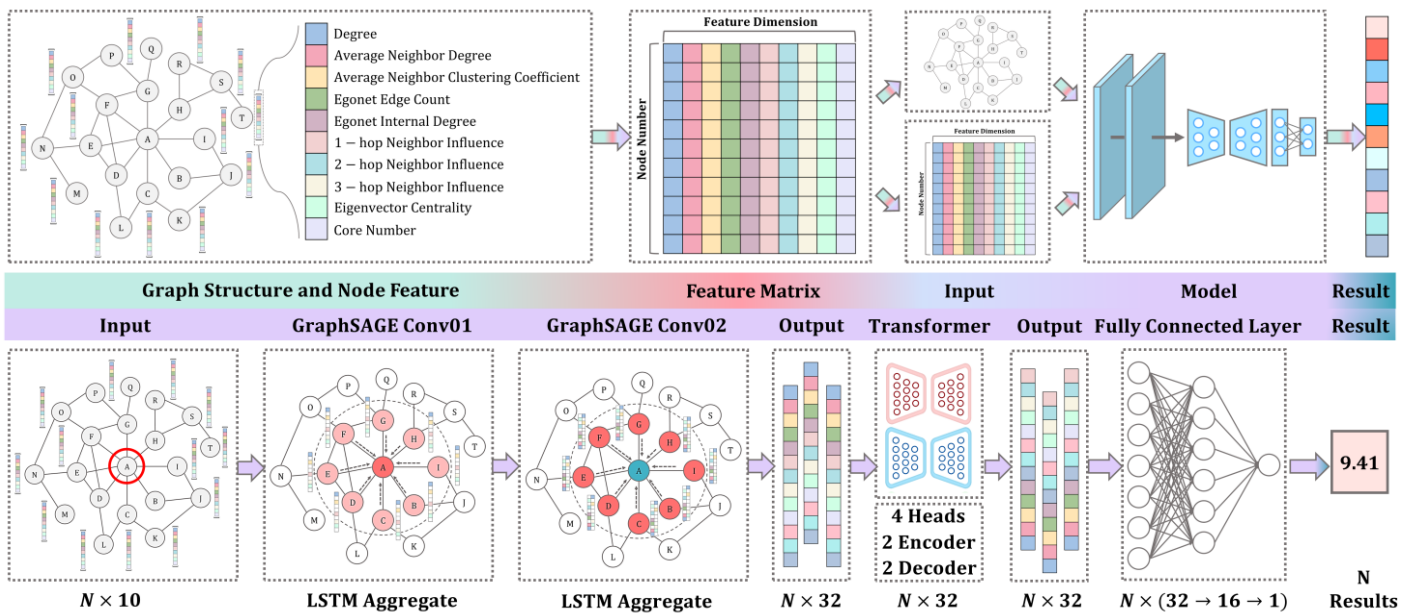

**Figure 02: The Structure of the GNNT Model** The structure of GNNT is divided into three parts: (1) two GraphSAGE convolutional layers using LSTM aggregators; (2) a Transformer network with two encoder and two decoder layers; and (3) fully connected layers for mapping the captured high-dimensional features to node influence values. The model's input consists of the network structure and a feature matrix, and the output is the predicted influence values of the nodes. Each node has 10-dimensional features; for detailed information, please refer to section A of Supplementary S02.

During the pre-training phase of the model, we constructed several small random networks and artificial networks and trained a predictor based on the SIR simulation results of these networks. We refer to the pre-trained base model as the GraphSAGE Neural Network with Transformer (GNNT). In the pre-training process, node labels are defined by the number of nodes infected by each node after the propagation process concludes. To mitigate the inherent randomness of the SIR propagation model, each node's label is the average result of 1,000 simulation experiments. The GNNT predictor consists of three parts: firstly, two GraphSAGE convolutional layers employing LSTM aggregators; secondly, a Transformer section comprising both encoder and decoder components; and finally, two linear layers that map the features captured by the model to the predicted influence values of the nodes. The structure of the model is illustrated in Figure 02, and further details can be found in the Methodology section and Supplementary S02.

After obtaining the pre-trained GNNT model, we employ the K-Means clustering method to cluster the nodes in real networks, identifying the core nodes of each cluster and combining them into a node set. Subsequently, we calculate the uncertainty of the model's predictions for these nodes and select the nodes with the highest uncertainty. We then fine-tune the GNNT on real networks using an active learning strategy based on clustering and uncertainty sampling, utilizing the labels of these nodes obtained from simulation experiments to further learn network-specific features. The fine-tuned model is referred to as the GraphSAGE Neural Network with Transformer using Active Learning (GNNTAL). Finally, we input the complete real network structure into the GNNTAL model to obtain the influence predictions for each node. For detailed information on the active learning methodology, please refer to the Methodology section and Supplementary S02.

# Results

## 1 Single Node Influence Prediction

### 1.1 Data

To evaluate the performance of GNNTAL, we conducted tests on twelve real-world networks. These networks were collected from various research domains, including biological networks (Dmela[45], C_elegans[45], Yeast[45,46]); transportation networks (USAir97[45], Minnesota[45]); social networks (Dolphins[45,47], Lesmis[45], Sport[45], Email[45,48], Jazz[45,49]); a citation network (NS_GC[45,50]); and a lexical network (Adjnoun[45,50]). The basic statistical characteristics of these networks can be found in Supplementary 09. The baseline methods used for comparison are Structural Neighborhood Centrality(SNC)[19], GLSTM[16], CycleRatio(CR)[51], RCNN[17], Local Information Dimensionality(LID)[31], KSGC[52], Social Capital(SC)[20], and H-index(HI)[21] (for detailed information on the baseline methods, please refer to Supplementary 01).

### 1.2 Accuracy of the GNNTAL

The SIR propagation model evaluation can effectively verify the accuracy of node influence ranking sequences[19]. We typically use the Kendall correlation coefficient to quantify its accuracy. Firstly, we conduct SIR simulation experiments for each node in the network to obtain and rank their propagation abilities[53]. We then calculate the Kendall correlation coefficient between the node influence sequence derived from the model and that obtained from the SIR experiments. The closer this coefficient is to 1, the more similar the model-generated sequence is to the SIR sequence, indicating higher accuracy of our method. To assess the performance of the GNNTAL model, we conducted comparative experiments on 12 real-world networks, with the results presented in Table 1. As shown in Table 1, the GNNTAL model demonstrates higher accuracy, achieving Kendall correlation coefficients above 0.9 in 6 out of the 12 networks. It ranked second in only two networks, with a minimal gap from the first place. Moreover, the active learning strategy based on K-Means clustering and uncertainty sampling enhanced the prediction capability of the base model. In all networks, the GNNTAL model fine-tuned through active learning outperformed the GNNT model in terms of accuracy. Indeed, the GNNT model, trained on random and artificially synthesized small networks, captures only some basic features present in the network.

Table 01: Kendall's Correlation Coefficient of Different Methods Results with SIR Simulation Results   The SIR simulation results for each network are the average outcomes of 1,000 independent experiments. A higher Kendall correlation coefficient indicates greater accuracy of the method.

| Network | GNNTAL | GNNT | SNC | GLSTM | CR | RCNN | LID | KSGC | SC | H-index |
|---|---|---|---|---|---|---|---|---|---|---|
| Adjnoun | **0.9306** | 0.8502 | 0.9133 | 0.8291 | 0.5803 | 0.6399 | 0.8569 | 0.9142 | 0.9012 | 0.8762 |
| Celegans | **0.8933** | 0.7062 | 0.8573 | 0.7128 | 0.4887 | 0.6939 | 0.7358 | 0.8465 | 0.8595 | 0.7841 |
| Dmela | **0.7104** | 0.6957 | 0.6909 | 0.5586 | 0.5411 | 0.5743 | 0.6365 | 0.6930 | 0.7012 | 0.6579 |
| Dolphins | **0.9222** | 0.8651 | 0.8852 | 0.7271 | 0.4903 | 0.5871 | 0.7820 | 0.8153 | 0.8638 | 0.8443 |
| Email | **0.9181** | 0.8498 | 0.87 | 0.7912 | 0.5328 | 0.6551 | 0.7872 | 0.8302 | 0.8932 | 0.8197 |
| Jazz | 0.9127 | 0.6481 | 0.8823 | 0.7770 | 0.5982 | 0.7139 | 0.8291 | 0.862 | **0.9140** | 0.8668 |
| Lesmis | 0.9146 | 0.8767 | **0.9363** | 0.7749 | 0.7197 | 0.7348 | 0.8653 | 0.8803 | 0.9233 | 0.8395 |
| Minnesota | **0.7621** | 0.5931 | 0.5381 | 0.2922 | 0.0024 | 0.4850 | 0.3956 | 0.6236 | 0.5245 | 0.3745 |
| NS_GC | **0.9109** | 0.8425 | 0.855 | 0.6094 | 0.2987 | 0.7157 | 0.6134 | 0.7801 | 0.8529 | 0.6133 |
| Sport | **0.6991** | 0.6740 | 0.6766 | 0.4402 | 0.3466 | 0.4635 | 0.5882 | 0.6769 | 0.6827 | 0.6072 |
| USAir97 | **0.8961** | 0.8732 | 0.8726 | 0.6964 | 0.3982 | 0.6729 | 0.7682 | 0.8395 | 0.8874 | 0.7943 |
| Yeast | **0.8561** | 0.7236 | 0.8462 | 0.6209 | 0.4276 | 0.6462 | 0.7607 | 0.8314 | 0.7932 | 0.7676 |

In addition to the Kendall correlation coefficient, this study also employs the Jaccard similarity coefficient to evaluate the ability of different methods to identify the Top-k nodes[54]. The Jaccard similarity coefficient measures the similarity between two sets by calculating the ratio of the intersection length to the union length of the sets. In this study, we calculate the Jaccard similarity coefficient between the top K elements of the node influence sequence derived from a given method and the top K elements of the node influence sequence obtained from the SIR simulations. The closer the Jaccard similarity coefficient is to 1, the higher the accuracy of the method. The Jaccard similarity coefficient results can be found in Supplementary 05, where GNNTAL consistently ranks among the top across all networks.

**1.3 Correlation Analysis**

Correlation analysis can help us understand and evaluate the relationship between two methods and their consistency or discrepancy in specific tasks[51]. In this study, we conducted a correlation analysis between the node influence sequences provided by the GNNTAL model and those derived from traditional methods (such as LID, CycleRatio, SNC and et al) . The results for the Yeast network are shown in Figure 2.

Figure 03(a) presents the correlation coefficients between the predictions of GNNTAL and those of traditional centrality methods on the Yeast network. Figures 03(b) and 03(c) show the correlations between GNNTAL's predictions and those of the deep learning models RCNN and GLSTM, respectively (for more results, see Supplementary 06). It can be observed that GNNTAL has a high degree of consistency with the SNC (0.88), KSGC (0.88), and SC (0.86) methods, indicating that

GNNTAL may have learned a similar approach to determining node influence as these methods. On the other hand, GNNTAL shows a certain degree of consistency with the LID (0.75) and HI (0.77) methods, though there are also some differences, which might be attributed to the varying sensitivities of different methods to distinct features in real networks.

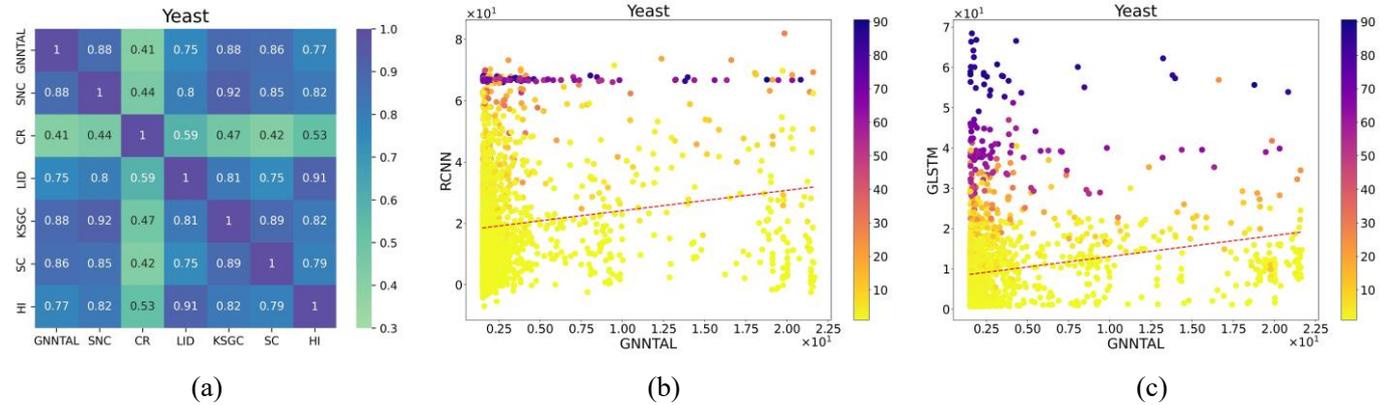

Figure 03: Correlation Analysis Results Between GNNTAL and Different Methods (a) Correlation analysis results between GNNTAL and traditional centrality methods. Each rectangle's value represents the Kendall correlation coefficient between the node influence sequences provided by the two corresponding methods. The darker the color, the more similar the node sequences given by the two methods. (b) Correlation analysis results between GNNTAL and RCNN. Each point in the figure represents a node, with the x-axis showing the node influence prediction values given by the GNNTAL model and the y-axis showing the node influence prediction values given by RCNN. The node color is defined by its SIR simulation results, with darker colors representing greater influence. (c) Correlation analysis results between GNNTAL and GLSTM.

Notably, in many networks, a high correlation is observed between GNNTAL and SNC, as well as between SNC and LID, but not as high between GNNTAL and LID. For example, in the Sport network, $Kendall(GNNTAL, SNC) = 0.80$, $Kendall(SNC, LID) = 0.81$, but $Kendall(GNNTAL, LID) = 0.64$; in the Celegans network, $Kendall(GNNTAL, SNC) = 0.84$, $Kendall(SNC, LID) = 0.83$, $Kendall(GNNTAL, LID) = 0.70$. Additionally, a similar situation is observed between GNNTAL and KSGC in the Yeast network. These phenomena suggest that GNNTAL does not solely rely on certain structural features to determine node influence, as traditional methods do, but rather considers a comprehensive range of global and local features.

Furthermore, as seen from Supplementary 06 and Figure 03, GNNTAL's node influence sequences show almost no correlation with those given by CycleRatio. This is because CycleRatio itself is not designed based on the SIR propagation model, nor does it use similar structural features as other methods to determine node influence.

For the two different deep learning models, GNNTAL shows almost no correlation with RCNN and GLSTM. Additionally, the wide dispersion of dark nodes (those with high SIR influence) in the figures indicates significant differences in the evaluation of high-influence nodes between GNNTAL and the RCNN and GLSTM models. Combining this with Figure 3(a), it is evident that although GNNTAL shows high consistency with methods like SNC and KSGC, its accuracy is higher. Therefore, GNNTAL can capture more complex network features through neural networks, thereby optimizing prediction results. In contrast, traditional centrality methods are typically based on predefined rules or formulas, lacking the capability to capture subtle differences and potentially high-level features present in the network.

Figure 04 illustrates the distribution of critical nodes identified by GNNTAL, LID, CycleRatio, and Social Capital in the Yeast network. In Figure 04, darker colors indicate higher node influence. It is evident from Figure 04 that, whether from the perspective of GNNTAL, LID, or SC, the critical nodes are concentrated in the core region of the network, exhibiting a significant "rich-club effect." In contrast, the critical nodes identified by CycleRatio are widely distributed throughout the network, with sparse connections among them. This distribution could potentially enhance the collective influence of these critical nodes, as their influence areas do not overlap significantly. For the GNNTAL model, its base model GNNT is ultimately trained based on the SIR simulation results of nodes. Therefore, the critical nodes identified by GNNTAL still inevitably exhibit clustering. To address the clustering issue, minor adjustments to GNNTAL's predictions are necessary in the influence maximization section to enhance its performance in solving the influence maximization problem.

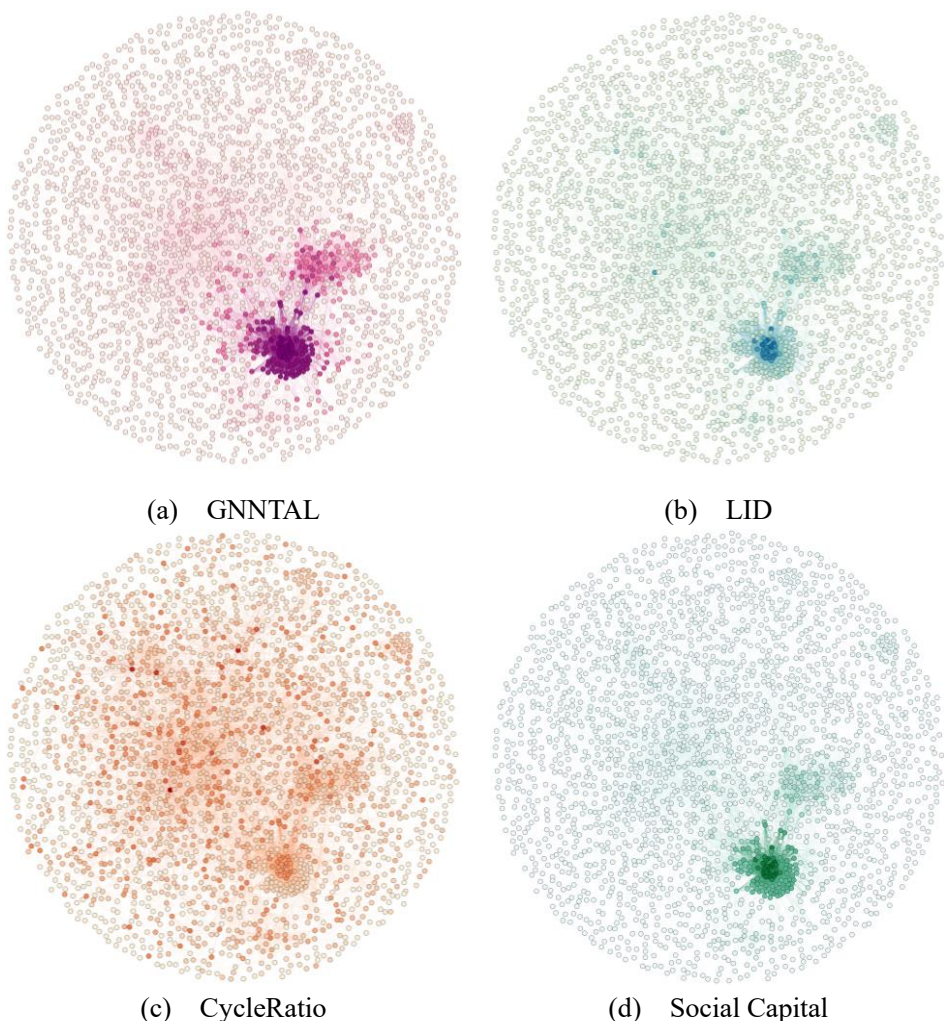

(a) GNNTAL  (b) LID
(c) CycleRatio  (d) Social Capital

**Figure 04:** Distribution of critical nodes identified by different methods in the Yeast network. (a-d) show the critical node distributions identified by GNNTAL, LID, CycleRatio, and Social Capital, respectively. In all four plots, the positions of the nodes and the network structure are identical. The color of each node represents its influence, with darker colors indicating higher influence.

## 2 Influence maximization problem

The influence maximization problem refers to the challenge of selecting a set of nodes in a network, under certain conditions, to maximize the spread of information under a given propagation model[16]. IMP is an NP-Hard problem[55], and existing methods typically rely on greedy algorithms or heuristic approaches to solve it[14]. Greedy algorithms offer high accuracy but often require evaluating all candidates at each step to select the current optimal solution, leading to redundant computations. This inefficiency makes them difficult to apply to large-scale networks. Heuristic algorithms, on the other hand, are prone to the "rich-club" phenomenon, where high-influence nodes cluster together. When these nodes are used as the seed set for propagation, their collective influence may decrease. To address this issue, some algorithms have considered the distribution of seed nodes within the network[28,56,57], aiming to select a dispersed set of nodes as the seed set. However, such methods require accurate estimation of node influence; otherwise, the influence of the selected nodes may still undermine each other.

In this study, we designed a simple greedy strategy on the node influence sequence predicted by the GNNTAL model to select seed nodes. This strategy, referred to as GNNTAL-Diversity constraints greedy Strategy (GNNTAL-DS), has a time complexity that typically does not exceed $O(N^2)$. In essence, we iterate through the node influence sequence, selecting the highest influence node with the fewest connections to the existing seed node set (for details, see the Methodology section and Supplementary 03), until all seed nodes are selected. To evaluate the performance of GNNTAL-DS, we conducted comparative experiments using both the Independent Cascade Model and the Linear Threshold Model. All experiments represent the average results of 1,000 independent trials. The baseline methods include CI[30], SNC, GLSTM, CR, RCNN, LID, KSGC, SC, and H-index (for detailed information on the baseline methods, please refer to Supplementary 01)

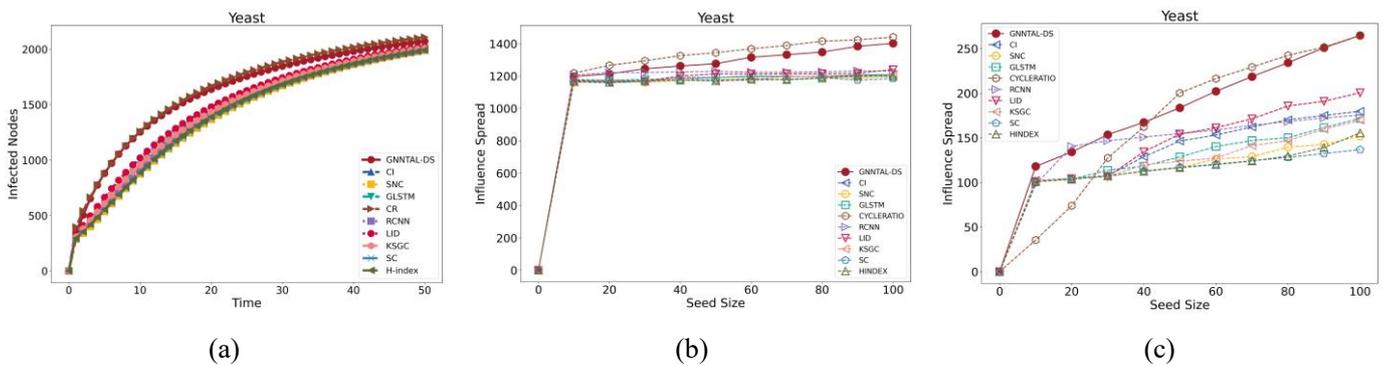

(a) (b) (c)

**Figure 05:** Influence of seed nodes selected by different methods in the Yeast network. (a) Rapid spread experiment results of different methods in the Yeast network using the SI model. The results represent the average of 1,000 independent experiments. In the figure, the x-axis represents time, and the y-axis represents the number of influenced nodes in the network at the current time. The higher the propagation curve, the better the performance of the method. (b) Maximum influence of seed nodes selected by different methods under the Linear Threshold Model. The x-axis represents seed size, and the y-axis represents the number of nodes ultimately influenced in the network. The greater the number of influenced nodes, the better the performance of the method. (c) Maximum influence of seed nodes selected by different methods under the Independent Cascade Model.

Figure 04 presents a comparison between GNNTAL-DS and the baseline methods on the Yeast network. The Collective Influence and GLSTM methods are specifically designed to solve the IMP problem. For the other baseline methods, the top-ranked nodes from their respective influence sequences are selected as the seed node set for comparison. As shown in Figure 04, GNNTAL-DS outperforms all baseline methods except CycleRatio in both the Independent Cascade Model and the Linear Threshold Model. In the Linear Threshold Model, GNNTAL-DS ranks second, with a slight gap behind CycleRatio. In the Independent Cascade Model, GNNTAL-DS achieves the best performance in some cases and slightly lags behind CycleRatio in others. Moreover, in the Linear Threshold Model, only GNNTAL-DS and CycleRatio show an increase in influence range as the seed set size increases, while the influence range for other baseline methods fluctuates around 1200 once the seed node count exceeds 20. Regarding propagation speed, GNNTAL-DS also surpasses all baseline methods except CycleRatio. GNNTAL-DS even leads CycleRatio during the initial phase of propagation, but shows weaker performance in the later stages, ultimately resulting in a disparity between GNNTAL-DS and CycleRatio.

As seen from Figures S07 and S08, in the Independent Cascade Model, GNNTAL-DS achieves an absolute advantage in four networks (Adjnoun, Lesmis, Dolphins, and NS_GC). It matches the performance of the best method in five networks (Jazz, USAir97, Email, C_elegans, and Yeast), but performs poorly in three networks (Sport, Minnesota, and Dmela). In the Linear Threshold Model, GNNTAL-DS gains an advantage in six networks and is weaker than CycleRatio in two networks, but still outperforms other methods. Overall, GNNTAL-DS performs better in the Linear Threshold Model than in the Independent Cascade Model, achieving high performance in small to medium-sized networks but performing poorly in large networks.

Why does GNNTAL-DS underperform in large-scale networks compared to other methods? We speculate that the following reasons might contribute:

(1)GNNTAL-DS is still based on node SIR propagation model simulations, which is an inherent limitation and restricts its generalization ability across different problems.

(2)The most crucial point is that our greedy strategy maximizes the assurance that node influence areas do not overlap. However, as the seed node set expands and the network size increases, the strategy tends to select low-influence nodes in peripheral areas to avoid overlap (this issue is discussed in detail in section B of Supplementary 03).

(3)Additionally, the mechanisms of different propagation models may also have an impact. For the IC model, a node has only one chance to infect other nodes, which emphasizes the propagation ability of individual nodes. In contrast, in the LT model, a node decides whether to be infected based on the state of its neighbors, highlighting the collective influence of a node's neighbors rather than the propagation ability of a single node. For both the LT and SI models, the infection process is continuous, emphasizing sustained propagation ability and the connectivity of the node's local network structure.

To address the aforementioned issues with GNNTAL-DS, the following approaches can be considered:

(1)Improving the node selection strategy. For instance, adjusting the selection strategy to prioritize influence over influence range (e.g., modifying the GNNTAL-DS selection strategy to choose nodes with number of connections less than or equal to $k$, but the appropriate $k$ value would need to be determined experimentally). Alternatively, employing more complex greedy algorithms that re-evaluate the influence range of each node after each selection to minimize overlap. While these methods can enhance accuracy, they may also increase computational complexity.

(2)Fine-tuning for different application scenarios to enhance the model's generalization ability.

(3)Incorporating CycleRatio results of nodes as features in the feature matrix and retraining the model.

## 5. Conclusion

Identifying critical nodes in complex networks is an age-old yet popular research topic. Traditional methods for identifying critical nodes in complex networks mainly focus on determining which structural features make nodes more influential. Unlike these traditional methods, this paper proposes a node influence prediction model, GNNTAL, based on GraphSAGE and Transformer. The model is initially pre-trained on random or artificial networks to enable it to capture general features. It is then fine-tuned on real networks using an active learning strategy that employs K-Means clustering and uncertainty sampling, allowing the model to capture unique features that may exist in real networks and enhancing its generalization ability. The combination of GraphSAGE and Transformer allows the model to thoroughly learn both local and global network features. Comparative experiments on 12 real-world network datasets demonstrate that GNNTAL achieves the best performance, closely approximating the results of the SIR model. This lays a solid foundation for future work.

For the IMP problem, we designed a simple greedy strategy, GNNTAL-DS, based on the prediction results of the GNNTAL model to select seed nodes. This strategy can achieve the performance of optimal methods with very low time complexity. We compared GNNTAL-DS with other baseline methods across all twelve networks. GNNTAL-DS performed better in the Linear Threshold Model than in the Independent Cascade Model, demonstrating high performance in small to medium-sized networks but performing poorly in large networks. Finally, this paper analyses the possible reasons for the poor performance of GNNTAL-DS in large networks and proposes improvement methods.

For future research, the approach of GNNTAL can be extended to more domains, such as higher-order networks or multilayer networks. There is no need to consider or design more complex structural features; modifications to the GNNT model would suffice. For example, GraphSAGE can be replaced with RGCN (suitable for heterogeneous graphs) to address the critical node identification problem in higher-order networks or heterogeneous networks. Alternatively, Transformer can be replaced with large models like BERT to solve the critical node identification problem in hypergraphs. Moreover, for other problems in complex networks, such as network control, we can fine-tune the GNNT based on its node influence predictions according to different task information and network structures. Additionally, we also could directly adjust the GNNT architecture design to meet the specific requirements of task selection based on different task information and network structures.

Certainly, GNNT still has its limitations. Firstly, GNNT is designed based on the SIR model, and the node influence sequences and scores provided by GNNTAL are intended to fit the SIR model's results as closely as possible. This limits GNNTAL's performance on other problems (e.g., network dismantling). Ideally, a general model should be trained and then fine-tuned for different tasks based on specific task information to enhance the model's generalization ability. This, of course, is left for future work. Secondly, the feature matrix used in this study is still manually set. GNNT employs a large number of features when constructing the node feature matrix, but whether these features achieve the expected results and whether there is overlapping influence between features require further analysis. Lastly, GNNT remains within the realm of supervised learning. Although GNNT significantly reduces the amount of data required for training, it may still require substantial annotation resources when dealing with ultra-large complex networks (the number of data samples required for the active learning strategy is at most 10% of the network nodes).

Nevertheless, deep learning methods still have unparalleled advantages over traditional methods. Deep learning is data-driven and does not rely entirely on network structure. We only need to input the network structure and feature matrix into the model and obtain the output data. Theoretically, as long as there is a network structure and corresponding feature matrix, deep learning models can solve the critical node identification problem for any complex network, especially higher-order or multilayer networks. In contrast, traditional methods require researchers to painstakingly design sophisticated structural feature methods. However, for deep learning models, we simply input the network structure and feature matrix, and let AI handle the rest. With the rapid development of artificial intelligence and large language models, we firmly believe that AI-based methods will gradually replace traditional heuristic algorithms.

## Methodology

### 1 Metrics:

**Kendall's Correlation Coefficient:** The Kendall correlation coefficient is a measure of the rank correlation between two sequences[21]. Given sequences A and B, the Kendall correlation coefficient $\tau$ between $A$ and $B$ is defined as follows:

$$\tau(A,B) = \frac{(N_{CP} - N_{DP})}{l(l-1)/2} \tag{1}$$

where $l$ is the number of nodes in the network, $N_{CP}$ denotes the number of concordant pairs, and $N_{DP}$ denotes the number of discordant pairs. Concordant pairs and discordant pairs refer to pairs of ordered pairs that have a certain relational sequence. For sequences $A$ and $B$, we take elements from each sequence to form pairs $(A_1, B_1), (A_2, B_2), \ldots$ We then create ordered pairs of these pairs, $((A_m, B_m), (A_n, B_n))$. If $A_m < B_m$ and $A_n < B_n$; or $A_m > B_m$ and $A_n > B_n$, they are termed concordant pairs. Otherwise, they are discordant pairs. In this study, we use the Kendall correlation coefficient to measure the consistency between the node influence sequences $M = (m_1, m_2, \ldots m_n)$ obtained by different methods and the node influence ranking sequence $S = (s_1, s_2, \ldots s_n)$ given by the standard SIR propagation model. If the Kendall correlation coefficient between a method's node influence sequence and the SIR model's node influence sequence is 1, it indicates that the method's sequence is identical to the sequence given by the SIR model. Notably, due to the inherent randomness of the SIR propagation model, the SIR node influence sequence used in this study is the average result of 1,000 independent experiments.

**Jaccard Similarity Coefficient:** The Jaccard Similarity Coefficient is commonly used to measure the similarity or difference between two sets[54]. Given two finite sets $U$ and $V$, the Jaccard similarity coefficient between these sets is defined as:

$$Jaccard(U,V) = \frac{|U \cap V|}{|U \cup V|} \tag{2}$$

In this study, we form finite sets by taking the top $N$ nodes from the node influence sequences obtained by baseline methods and compare them with the finite sets formed by the top $N$ nodes from the node influence sequence given by the SIR model. If the Jaccard similarity coefficient between the two sets is 1, it indicates that the top influential nodes identified by the baseline method are identical to those identified by the SIR model.

### 2 Model:

**Label:** Each node's label is defined by the results of the node's SIR Propagation Model Simulation. The SIR (Susceptible, Infected, Recovered) model is a standard epidemic dynamics model. In the SIR model, each individual's state can be one of three (S, I, R), and each individual can only be in one state at each time step. When an individual is in the susceptible state (S), it can change its neighboring nodes to the infected state (I) with a certain probability. Nodes in the infected state will transition to the recovered state (R) in the next time step and will no longer change states thereafter. In this study, we set each node in the network to the infected state one by one and perform simulations. The experiment ends when there are no infected nodes in the network. At this point, we count the number of nodes in the R state and consider this as the node's influence. For the probability of a node becoming infected (i.e., the transmission probability) $\beta_{th}$, we set it as $\beta_{th} = 1.01 \times \beta$ where $\beta$ is the transmission threshold of the SIR model, typically set as[58,59]:

$$\beta = \frac{<k>}{<k^2> - <k>} \tag{3}$$

where $<k>$ is the average degree of the network, and $<k^2>$ is the mean square of the network node degrees.

**GNNT:** The GNNT model is a node influence prediction model pre-trained on random and artificial networks. The model takes as input the network structure and node features (10 dimensions) and outputs the predicted influence values of the nodes. The GNNT model consists of three parts. First is the GraphSAGE component, which primarily extracts local features of the nodes in the network. This is followed by the Transformer component, which captures potential long-range dependencies between nodes, especially global features of the network. After the Transformer, there are two fully connected layers that map the high-dimensional node features to the target dimension, ultimately outputting the model's predictions. Specifically, the GraphSAGE component comprises two SAGE convolutional layers. The input dimension of the first convolutional layer is $N \times 10$, and the output dimension is $N \times 32$. The second convolutional layer has an input and output dimension of $N \times 32$. Each convolutional layer samples all first-order neighbor nodes, and the aggregation method used is LSTM aggregation. This sampling-aggregation method captures more detailed and complex local features. The Transformer component consists of two encoders and decoders. Each encoder includes two linear layers (serving as feedforward networks) and two normalization layers (with one following the multi-head attention mechanism and another following the feedforward network). Each decoder includes two linear layers (as feedforward networks) and three normalization layers (with one following the multi-head self attention mechanisms, one following the multi-head attention mechanisms and another following the feedforward network). Additionally, the multi-head attention mechanism in the Transformer has four heads, and the dropout rate is 0.1. The fully connected layer component consists of two linear layers. The first fully connected layer maps the data from 32 dimensions to 16 dimensions, and the second fully connected layer maps the data from 16 dimensions to 1 dimension, which is the final result. For more detailed information on the model, refer to Supplementary 02.

**Active Learning:** The advent of active learning addresses the issue of high labeling costs and the difficulty of obtaining large amounts of labeled data in supervised learning[60,61]. Taking the node influence prediction problem as an example, it typically requires the number of nodes each node infects in SIR simulations, averaged over 1,000 independent experiments, as labels. This is nearly impossible to achieve for large-scale real networks. Active learning can mitigate this by selecting a small number of representative nodes using a manually designed sample selection strategy, obtaining labels through SIR simulations, and fine-tuning the pre-trained model, thereby making it applicable to various types of real networks. The sample selection strategy set in this paper is based on an uncertainty sampling strategy with K-Means clustering, as follows: first, the K-Means algorithm is used to cluster the node features in the network; then, core points (the points closest to the cluster center) are selected from each cluster as representatives and added to the node set; thereafter, we select the top N nodes with the highest uncertainty from the node set (N essentially corresponds to the number of feature clusters. For small networks, N is usually set to 50, and for medium to large networks, N is typically $\geq 200$, but it will not exceed 10% of the total number of nodes. For details on N selection, see section C of Supplementary 02) as samples for active learning. The uncertainty of a node is defined by the variance of its prediction results. Through clustering and uncertainty sampling, we can ensure that the sampled data covers the entire feature space as much as possible, guiding the model to focus on learning difficult-to-predict samples, thereby enhancing the model's generalization ability and prediction accuracy.

## 3 Influence Maximization Problem

**Node Selection Strategy:** The IMP problem is typically addressed using heuristic algorithms or greedy algorithms. However, important nodes identified by traditional heuristic algorithms often exhibit significant clustering, leading to overlapping influence areas and consequently weakening the collective influence of the node set. In this study, we derive the seed node set using a simple greedy strategy based on the node influence sequence generated by the GNNTAL model. The fine-tuning strategy is as follows: First, we obtain the node influence sequence from the GNNTAL model, which closely aligns with the node influence sequence produced by the SIR propagation model. Next, we traverse this sequence, sequentially adding nodes to the seed node set. For each node, we calculate its number of connections with the already selected nodes. If the node has zero connections with the selected nodes, it is added to the seed node set. If not enough seed nodes are selected after one round of traversal, we continue to select the highest influence nodes that have not been chosen yet until all seed nodes are selected. The number of connections for a node is defined by its neighboring nodes. We simply count how many of the node's neighbours have already been selected as seed nodes to determine its number of connections. In short, the number of connections for a node is the number of seed nodes within its neighborhood. For detailed information on the node selection strategy, please refer to Supplementary 03.

**Independent Cascade Model:** The Independent Cascade Model is a type of network information propagation model. In a network, if there is an edge between two points, we can say node $a$ can influence node $b$. In a directed graph, the influence of node $a$ on node $b$ may not be the same as the influence of node $b$ on node $a$. In the Independent Cascade Model, given any set of seed nodes, the nodes in the set will activate their neighbouring nodes with a certain probability. The newly activated neighbouring nodes will, in turn, attempt to activate their remaining inactive neighbors, and this process continues until no more nodes can be activated in the network. The most notable feature of the Independent Cascade Model is that each activated node has only one chance to activate its neighbors, and each node's activation attempt is independent of others. In this study, the probability of a node activating its neighbors is denoted as $\beta_{th}$.

**Linear Threshold Model**：The Linear Threshold Model is another type of network information propagation model. In the Linear Threshold Model, each edge is assigned a weight $w$, which describes the proportion of the total influence that node $a$ has on node $b$. The larger the value of $w$, the greater the influence of node $a$ on node $b$. Additionally, each node has an influence threshold $\theta$. A higher $\theta$ indicates that the node is more difficult to influence, whereas a lower $\theta$ means the node is easier to influence. The key difference between the Linear Threshold Model and the Independent Cascade Model is that, in the LT model, each node has multiple opportunities to activate its neighboring nodes, whereas in the IC model, each node has only one chance to activate its neighbors.

## Data availability

All data that support the findings of this study are included within the article (and any supplementary files).

## Code availability

The custom code that supports the findings of this study is available at the following github repository: https://github.com/WHNetwork/GNNTAL

## Acknowledgement

This work was funded by the Key Research and Development Program of Yunnan Province (202102AA100021); the National Natural Science Foundation of China (62066048 and 62366057); the demonstration project of comprehensive government management and large-scale industrial application of the major special project of CHEOS: 89-Y50G31-9001-22/23; and the Science Foundation of Yunnan Province (202101AT070167) and supported by a grant from the Key Laboratory for Crop Production and Smart Agriculture of Yunnan Province (2022ZHNY10).

## Author contributions

H.W. and T.L. conceived the idea and designed the experiments; H.W. wrote this paper. T.L., S.Y., and M.J. developed the code. N.Z. provided effective guidance and suggestions. J.W. and N.Z. provided funding. J.W. and N.Z. revised the paper. All authors discussed the results and reviewed the paper.

## Competing interests

The authors declare no competing interests.